\documentclass[proceedings]{JHEP3} 
\PrHEP{hep2001}
\conference{International Europhysics Conference on High Energy Physics} 
\title{Theoretical progress in describing the $B$-meson lifetimes}
\author{{DAMIR BECIREVIC} \\ 
Dipartimento di Fisica, Universit\`a di Roma ``La
Sapienza'',\\ Piazzale Aldo Moro 2, I-00185 Roma,
Italy \\
E-mail: \email{Damir.Becirevic@roma1.infn.it}}

\usepackage{epsfig}

\def \beq{\begin{equation}}
\def \eeq{\end{equation}}
\def \bea{\begin{eqnarray}}
\def \eea{\end{eqnarray}}
\def \ben{\begin{enumerate}}
\def \een{\end{enumerate}}
\def \bit{\begin{itemize}}
\def \eit{\end{itemize}}

\abstract{
The present status of the theoretical estimates of the difference between the
widths of the neutral $B_s$-mesons and of the $B$-meson lifetime ratios is
reviewed. 
In particular, the lattice results for the matrix elements of  
the relevant $\Delta B=2$ operators are updated and the first 
lattice QCD results for the matrix elements of $\Delta B=0$ operators
are presented. 
In both cases, the NLO perturbative QCD corrections in the coefficient 
functions have been included. The theoretically updated results are:
$(\Delta \Gamma/\Gamma)_{B_{s}} = 7 \pm 4\ \%$, 
${\tau(B^+)/ \tau(B_{d})} = 1.07(3)$ and 
${\tau(B_s)/ \tau(B_{d})} = 1.00(2)$. }

\begin{document}
To make it clear and simple, I split the discussion into two parts: 
\begin{itemize}
\item[$\circ$] {\underline{$(\Delta \Gamma/\Gamma)_{B_{s}}$}}, the quantity 
that recently attracted quite a bit of attention among theorists and for which
the experimental upper limit has been set at~\cite{lep+cdf}:
\bea
\left({\Delta \Gamma \over \Gamma}\right)_{B_{s}} < 0.31\quad (95\%\ {\rm C.L.})\,.
\eea
\item[$\circ$] {\underline{$\tau(B_{u (s)})/\tau(B_{d})$}}  
have been measured quite accurately~\cite{lep+cdf}
\bea
{\tau(B_{u})\over \tau(B_{d})} = 1.07(2)\,,\quad {\tau(B_{s})\over \tau(B_{d})} =
0.95(4) \;.
\eea
Important theoretical progress in computing these ratios has been 
made this year. I will not discuss the theoretical predictions 
for the ratio  ${\tau(\Lambda_b)/\tau(B_{d})}$, 
where, in my opinion, substantial progress is yet to be made.
\end{itemize}

Theoretical set-up for both of the above topics relies on the hypothesis 
of the (global and local) quark--hadron duality~\cite{shifman}. 
The validity of that assumption is not totally clear, although the impressive 
agreement of many theoretical predictions in 
$\tau$-physics (for which the duality has been assumed) with the
precise experimental data is very 
encouraging~\cite{pich}.

\section{WIDTH DIFFERENCE OF THE $B_{S}^0$-SYSTEM}

The Orsay group broke the duality problem a little bit open by demonstrating 
that in the combined $N_c\to \infty$ and SV limit~\footnote{SV (Shifman--Voloshin limit) 
is the limit in which $\Lambda_{\rm QCD} \ll m_b - 2 m_c \ll m_b$~\cite{SV}.}, 
the quark--hadron duality  for $\Delta \Gamma_{B_s}$ indeed works~\cite{orsay}. 
They actually showed that the two channels, $B_s^0 \to D_s \overline D_s$, 
$D_s^\ast \overline D_s^\ast$ (S-wave), saturate 
the partonic expression for $\Delta \Gamma_{B_s}$. Out of that  
limit, however, the quark--hadron duality is again an assumption.

\FIGURE{
\epsfig{file=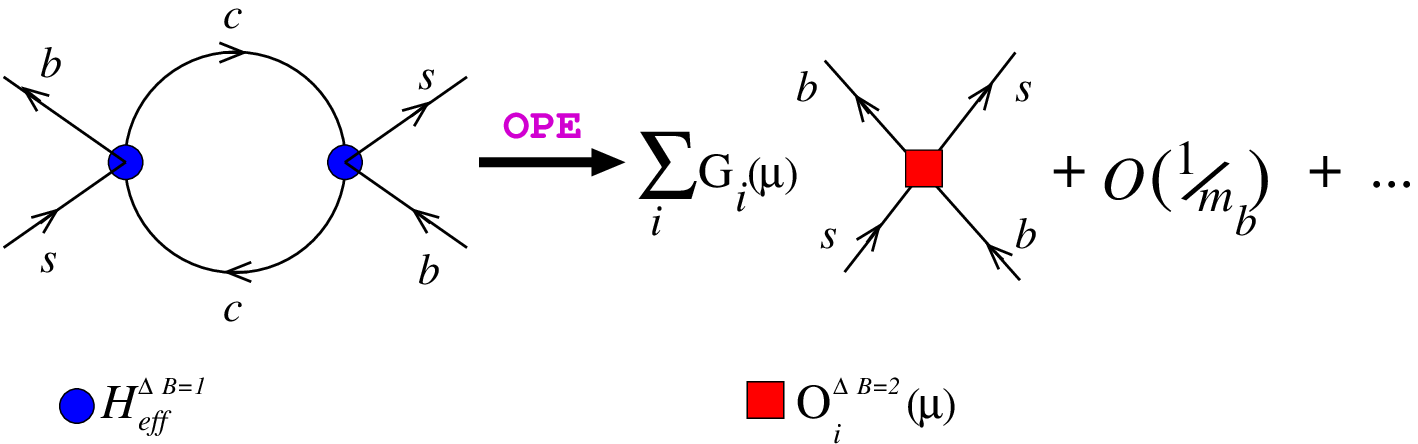,width=.8\linewidth}
\caption{\sl Heavy quark expansion: the non-local T-product of the l.h.s. 
(with the doubly inserted ${\cal H}_{eff}^{\Delta B=1}$) is expanded in 
the series in $1/m_b$, each coefficient being the sum of local 
${\Delta B=2}$ operators.}
\label{fig1}} 
The (modern) theoretical expression for 
$\Delta \Gamma_{B_{s}}$ has been 
derived in ref.~\cite{dunietz}, where the  operator product 
expansion (OPE) has been applied to compute the absorptive amplitude for the 
$B_s \to \bar B_s$ transition.~\footnote{For recent 
reviews on the computation of $(\Delta \Gamma/\Gamma)_{B_{s}}$, see also refs.~\cite{lenz,nierste}.} The high energy scale 
is provided by the inverse $b$-quark mass, which is why this expansion is
usually referred to as the heavy quark expansion (sketched in fig.~\ref{fig1}). 
The final expression of ref.~\cite{dunietz} looks as follows:
\bea \label{eq3}
\Delta \Gamma_{B_s} = {G_F^2 m_b^2\over 12 \pi m_{B_s}} \vert V_{cb}^\ast V_{cs}
\vert^2 \ \biggl\{ G_1(\mu) \langle \bar B_s\vert O_1(\mu)\vert  B_s\rangle 
+  G_2(\mu) \langle \bar B_s\vert O_2(\mu)\vert  B_s\rangle + \delta_{1/m_b}
\biggr\}\ ,
\eea
where the flavour structure of the operators $O_{1,2}(\mu)$ is $\Delta B=2$; 
$\delta_{1/m_b}$ contains all the contributions from 
the $1/m_b$ corrections to the first two terms. Corrections $\propto 1/m_b^n$
($n\geq 2$) are neglected.
\begin{itemize}
\item[$\clubsuit$] Short distance physics is encoded in the functions   
$G_{1,2}(\mu)$ which are the combinations 
of $\Delta B=1$ Wilson coefficients. The next-to-leading order (NLO)  
corrections to these functions have been computed in ref.~\cite{beneke}, 
where the authors also kept the ratio $m_c/m_b$ different from zero. Of  
conceptual importance is the fact that they explicitly 
verified the infrared safety of the functions $G_{1,2}^{\rm NLO}(\mu)$, as anticipated 
years before, in ref.~\cite{uraltsev}.  
Phenomenologically, however, the (subleading) corrections are uncomfortably large. 
For example, the dominant term changes as 
\bea
{  G_2^{\rm NLO} (m_b) - G_2^{\rm LO}(m_b)\over G_2^{\rm LO}(m_b) } \simeq -
0.35\; ,
\eea
{\it i.e.} the radiative corrections lower $G_2(m_b)$ by $35\% $. 
The residual scale dependence of $\Delta B=1$ Wilson coefficients entering 
the functions $G_{1,2}(m_b)$ is customarily 
estimated by varying the renormalization scale from 
$\mu=m_b/2$ to $\mu = 2 m_b$, which amounts to an error 
of $-20\%$ and $+15\%$, respectively;
\item[$\clubsuit$] Long distance QCD dynamics is described by the matrix 
elements, which are parametrized as
\bea \label{eq1}
&&\hspace*{-7mm}\langle  \bar B_s\vert O_1(\mu)\vert  B_s\rangle \equiv \langle  \bar B_s\vert (\bar b s)_{V-A}
 (\bar b s)_{V-A} \vert  B_s\rangle = {8\over 3} f_{B_s}^2 m_{B_s}^2 B_1
 (\mu)\,,\cr
&& \cr
&&\hspace*{-7mm}\langle  \bar B_s\vert O_2(\mu)\vert  B_s\rangle \equiv \langle  \bar B_s\vert (\bar b s)_{S-P}
 (\bar b s)_{S-P} \vert  B_s\rangle = -{5\over 3} \left({f_{B_s} m_{B_s}^2
 \over m_b(\mu)+ m_s(\mu) } \right)^2  B_2(\mu)
 \,,\cr
&& \cr
&& \hspace*{-7mm}\langle \bar B_s\vert O_3(\mu)\vert  B_s\rangle \equiv \langle  \bar B_s\vert (\bar b^i s^j)_{S-P}
 (\bar b^j s^i)_{S-P} \vert  B_s\rangle = {1\over 3} \left({f_{B_s} m_{B_s}^2
 \over m_b(\mu)+ m_s(\mu) } \right)^2  B_3(\mu)
 \,, 
\eea
where the third operator has the same Dirac structure as $O_2$ but with 
reversed colour indices ($i,j$), and hence mixes with $O_2$ under 
renormalization. 
The above $B$-parameters are all equal to unity in the vacuum saturation
approximation (VSA). A priori, VSA gives a gross estimate and one has to  
include the (non-factorizable) non-perturbative QCD effects. 
QCD simulations on the lattice represent a suitable 
method for that part of the job, which I discuss in the next subsection.
\end{itemize}

\subsection{$B$-parameters (novelties from the lattice)}

I would like to stress that, in principle, 
lattice QCD approach allows the fully non-perturbative estimate
of the hadronic quantities to an arbitrary accuracy. In practice,
however, many approximations need to be made which, besides the statistical, 
introduce various systematic uncertainties in the final results. 
The steady progress in 
increasing the computational power, combined with various 
theoretical improvements, helps reducing ever more of those  
systematic uncertainties. 
This is why the lattice QCD approach is so attractive.

The ultimate goal in the study of the heavy quark physics on the lattice is 
to produce the results by simulating the $b$-quark directly, in the full QCD. 
Since we are still quite far from that point, as I will briefly explain 
in what follows, we use various ways to treat the heavy quark  on the lattice 
and thus various ways to compute the $B$-parameters of eq.~(\ref{eq1}):
\begin{itemize}
\item[$\otimes$] {\underline{HQET}}: After discretizing the 
HQET lagrangian (to make it tractable for a lattice study), 
the matrix elements from eq.~(\ref{eq1}) were computed 
in ref.~\cite{GR}, but only in the static limit ($m_b \to \infty$)~\footnote{
In these
effective approaches (HQET, NRQCD), the light quark is, of course, treated relativistically ({\it i.e.}  
by using the standard (Wilson) QCD action).}.  
\item[$\otimes$] {\underline{NRQCD}}: A step beyond the static 
limit has been made in ref.~\cite{hiroshima}, where  
the $1/m_b$-corrections to the NRQCD lagrangian have been included, as well as  
a large part of $1/m_b$-corrections to the  
matrix elements of the four-fermion operators. 
A grain of salt, however, comes with the non-existence 
of the continuum limit of NRQCD on the lattice, so that one should find  
a window in which the discretization effects are simultaneously small for both,
the light quark ${\cal O}(a m_q)$ and the heavy one ${\cal O}(1/(a m_b))$~\footnote{
$a$ is the finite lattice spacing, which the authors chose to be close to $1/a \simeq 2$~GeV.}. 
\item[$\otimes$] {\underline{QCD}}: 
It would be preferable to treat the $b$-quark relativistically too, 
but such a study requires a huge computational power ({\it i.e.} very fine lattices 
to resolve a tiny $b$-quark wavelength), which is well 
beyond the capabilities of currently available parallel computers. 
For that reason, in ref.~\cite{ape}, the matrix elements were computed 
in the region of masses close to the charm quark and then extrapolated 
to the $b$-quark sector by using the heavy quark scaling laws (HQSL).  
This extrapolation, however, is very long and introduces 
large systematic 
uncertainty.
\end{itemize}
As of now, none of the above approaches is good enough on its own and all of them 
should be used to check the consistency of the obtained results.

This year progress in reducing   
the systematics of the heavy quark extrapolation of 
the $B$-parameters~(\ref{eq1}) 
has been reported in ref.~\cite{damir}. 
Besides several `minor' (albeit important) improvements,  
we combined the static HQET results of ref.~\cite{GR} with those of 
ref.~\cite{ape}, where 
lattice QCD is employed for three mesons of masses, $1.8\ {\rm GeV} 
\lesssim m_{P} \lesssim 2.4$~GeV. 
To use the HQSL we matched the QCD matrix elements 
with the HQET ones,~\footnote{
HQET is built on the heavy quark symmetry so that the 
HQSL are the manifest features of the theory. According to HQSL 
our  $B$-parameters should scale with the 
inverse heavy quark (meson) mass as a constant. The symmetry breaking terms 
are $\propto 1/m_Q^n$, and can be studied from our lattice data (we compute 
$B$-parameters for several fixed values of $m_Q$). } 
so that we could actually ``interpolate'' to the mass of the $B_s$-meson.  
The obtained results are then matched back to their full QCD values. This 
matching ${\rm HQET}\leftrightarrow {\rm QCD}$ ($C_B(m_P)$ in fig.~\ref{fig2}) 
is for the first time made at NLO in perturbation theory.

There is a point concerning the renormalization schemes that might
look messy, which I would like to explain here. 
The $\overline{\rm MS}$(NDR) schemes are unambiguously specified only at NLO. 
For consistency, we need to compute the $B$-parameters precisely in the 
$\overline{\rm MS}$(NDR) scheme used to compute the functions $G_i(\mu)$~\cite{beneke}.

{\bf i)}Operators computed in the static limit of HQET on the lattice 
have been matched onto the continuum ones at NLO (and thence renormalized in 
a well determined $\overline{\rm MS}$(NDR) scheme) by using the expressions derived in 
(boosted) perturbation theory~\cite{GR1}. 
The two-loop anomalous dimension matrices 
for all $\Delta B=2$ operators in HQET were computed in ref.~\cite{GR2}, 
so that their evolution and matching to/from the QCD operators, 
renormalized in the $\overline{\rm MS}$(NDR) scheme of ref.~\cite{beneke},
can be made unambiguously (at NLO).

{\bf ii)} In lattice QCD, the operators are non-perturbatively renormalized 
in the so-called (Landau)RI/MOM renormalization scheme and then converted to the 
$\overline{\rm MS}$(NDR) scheme of ref.~\cite{beneke} by using  
the NLO conversion formulae.  
\FIGURE{
\epsfig{file=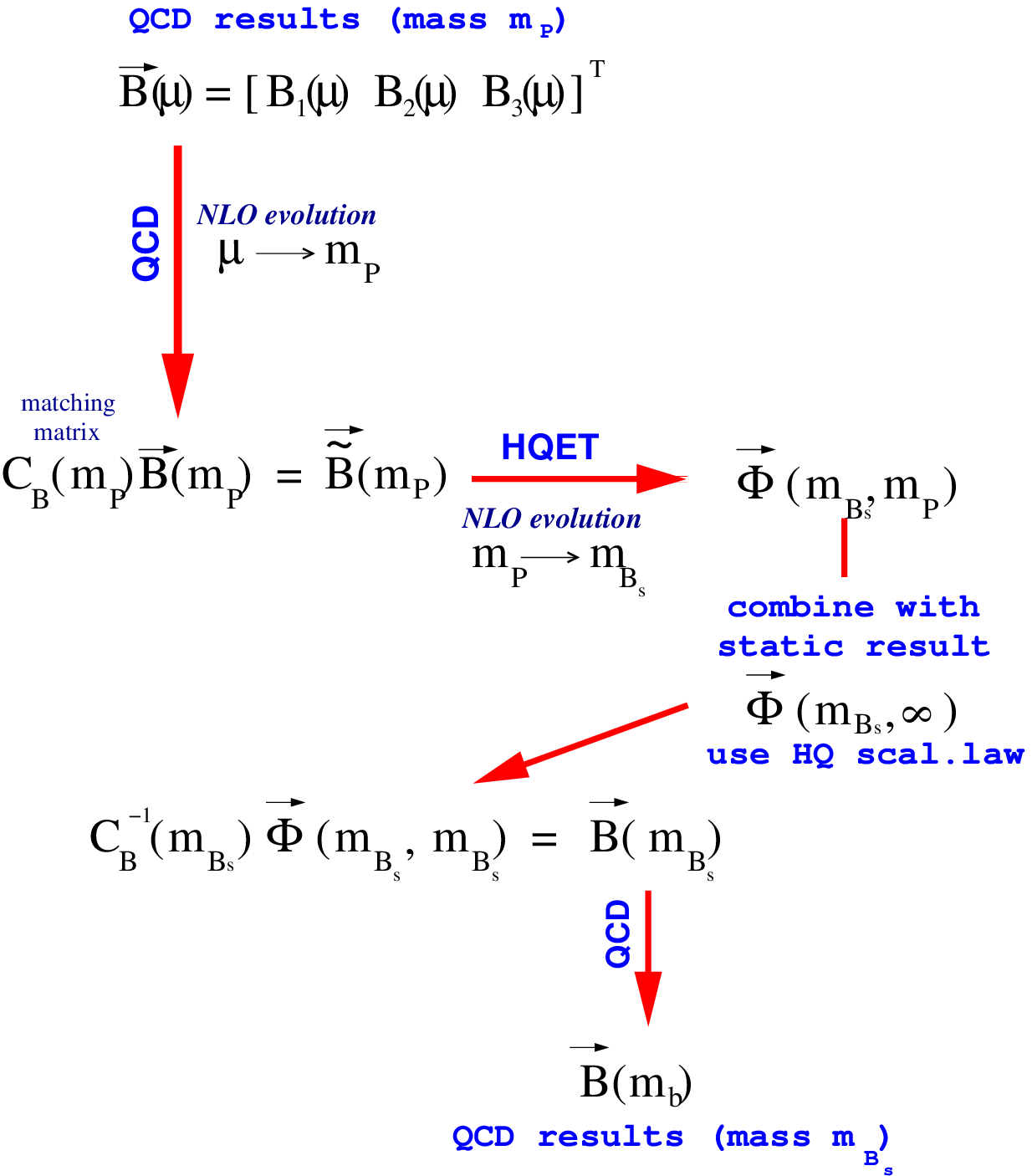,width=0.49\linewidth}
\epsfig{file=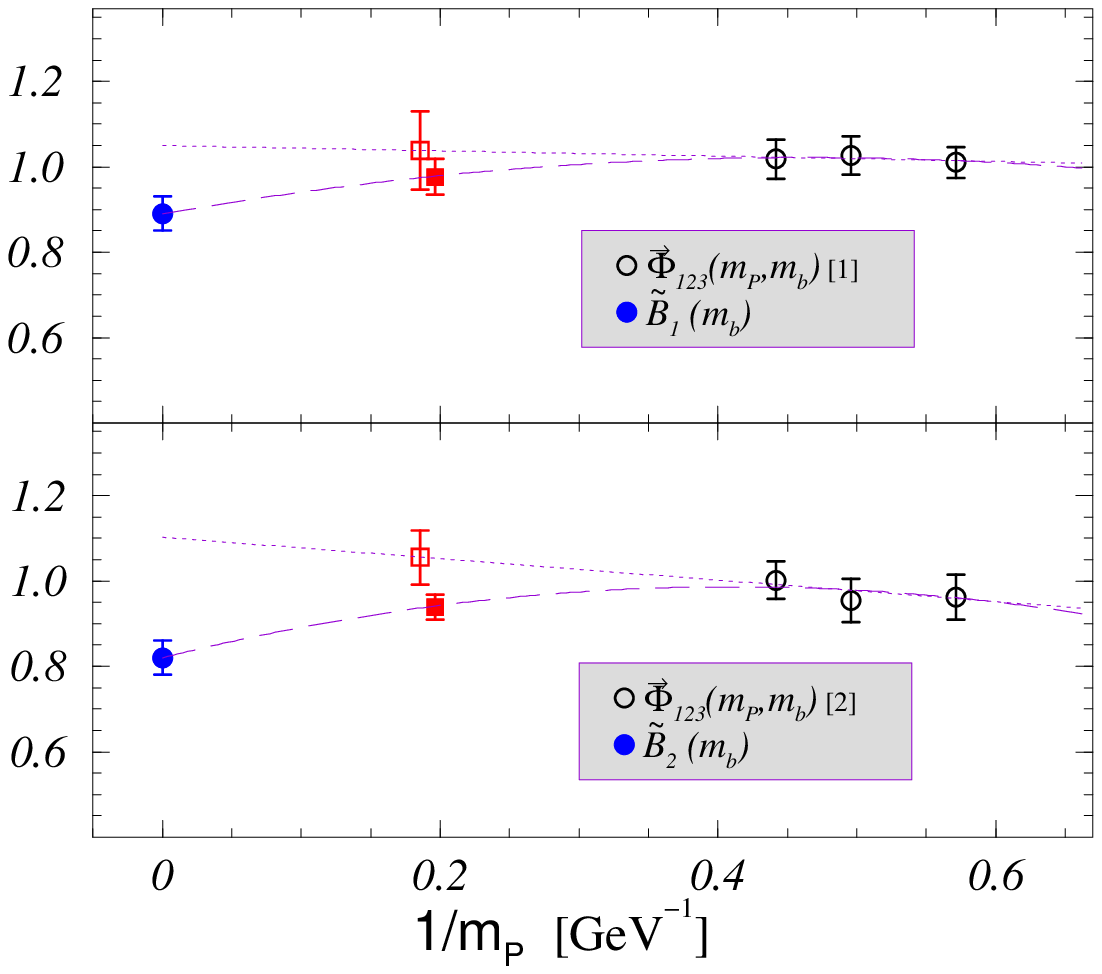,width=0.49\linewidth}
\caption{\sl The procedure of matching $QCD \leftrightarrow HQET$ is sketched and 
the mass `interpolation' illustrated. We combine the results 
obtained in QCD with three heavy--light mesons $m_P$, with the static HQET result, 
$m_P \to \infty$. The result of the linear extrapolation to $1/m_{B_s}$ is marked by 
the empty 
squares, whereas the ``interpolation'' is denoted by the 
filled squares. [1] and [2] are the components of the vector $\vec \Phi$, 
which are, after getting to $1/m_{B_s}$, matched back to the QCD parameters 
$B_{1,2}(m_b)$, the results of which are given in eq.~(\ref{mi}).}
\label{fig2}} 

{\bf i)} and {\bf ii)} ensure that the final results for $B$-parameters are 
indeed the ones corresponding to the $\overline{\rm MS}$(NDR) scheme of 
ref.~\cite{beneke} in QCD.
The schematic procedure of matching  
and the ``interpolation'' to $m_{B_s}$, are shown 
in fig.~\ref{fig2}. 
We obtain the following results
\bea \label{mi}
 B_1 (m_b) = 0.87(2)(5)\;, \hspace*{8mm}
 B_2 (m_b) = 0.84(2)(4)\;,
\eea
where the first errors are statistical and the second include various sources 
of systematics. An important remark is that the above results 
are obtained in the quenched approximation ($n_f=0$), and the systematic 
error due to quenching could not be estimated. This year's novelty is 
the research made in that direction by the JLQCD 
collaboration~\cite{norikazu}. Within the NRQCD approach, they examined 
the effect of inclusion of the dynamical quarks. They conclude that the 
$B$-parameters are essentially insensitive to switching from $n_f=0$ to 
$n_f = 2$. 
From their (high statistics) unquenched simulation, they quote
\bea \label{oni}
 B_1 (m_b) = 0.87(4)(7)\;,\hspace*{8mm} B_2 (m_b) = 0.86(3)(7)\;.
\eea
Notice that the two new lattice results (eqs.~(\ref{mi}) and (\ref{oni})) 
are in very good agreement.

\subsection{Phenomenology: taking all pieces together}
We can now either follow ref.~\cite{beneke} and write
\bea \label{bbgln}
&&\left({\Delta \Gamma \over \Gamma}\right)_{B_{s}} = \Bigl(
5.03(32)\ {\rm GeV}^{-2} \Bigr) \cdot 
\; f_{B_s}^2 B_1(m_b)\;  
{\cal M}(m_b)\;,\\
&&\hspace*{-32mm}{\rm where\;} \; \  {\cal M}(m_b) = G_1 (m_b) - G_2 (m_b)  {
\langle \bar B_s\vert O_2(m_b)\vert  B_s\rangle \over 
\langle \bar B_s\vert O_1(m_b)\vert  B_s\rangle } + \delta_{1/m} \;,\nonumber
\eea
or, as proposed in~\cite{ape}, we can write  
\bea \label{ours}
\left({\Delta \Gamma \over \Gamma}\right)_{B_{s}} = \Bigl(1.09(10)\times 10^{-5} \Bigr) \cdot 
{\,\left( \tau_{B_s} \Delta m_d \right)^{\rm EXP.}   \over \left| {V_{tb}  V_{td}} \right|^2 } 
\, \xi^2 \, {\cal M}(m_b)\;.
\eea
Indeed, by using $\xi^2$ 
and the experimental value for  
$\Delta m_d = 0.519(20)(16)\ \mbox{ps}^{-1}$~\cite{babar}, 
we avoid the multiplication by $f_{B_s}$, for which the uncertainty is much 
larger than the one for the ratio 
$\xi = f_{B_s} \sqrt{B_1^{(s)}}/f_{B_d}\sqrt{ B_1^{(d)}} = 
1.15(6)$~\cite{bernard}, 
for which many of the systematic errors cancel.
The critique has it that since the value
for $\vert V_{tb}  V_{td}\vert $ is needed to evaluate eq.~(\ref{ours}), 
one has to have recourse to their values obtained from the unitarity triangle 
analysis~\cite{ciuchini}, which implies that we assume the validity of the Standard Model 
(SM)~\cite{lenz}. I believe, however, that the assumed quark--hadron duality
is more of an issue than the validity of the SM, and therefore we should 
rather focus our attention on testing the duality (hypothesis) 
{\it within} the SM (theory).

Moreover, 
I do not see sense in looking for the physics beyond the SM from this 
quantity before taming the $1/m_b$ corrections. To back this claim, let us 
use the parameters~(\ref{mi}) and write the contributions to eq.~(\ref{ours}) 
term by term
\bea \label{num}
\left({\Delta \Gamma \over \Gamma}\right)_{B_{s}} = 0.005(9)\ +\ 
0.146(28)\ -\ 0.086(19) =  6.5\pm 2.0^{ +1.3}_{-2.1} = 6.5^{ +2.4}_{-2.9}\ \%\;,
\eea
where I also used $\delta_{1/m} = -0.5(1)$, 
as it can be obtained by applying 
the VSA to estimate the values of all the matrix elements that contribute 
at $1/m_b$ (identified in 
ref.~\cite{dunietz}). The error in $\delta_{1/m}$ is an {\sl ad hoc}
estimate. 
Note that in~(\ref{num}) I added separately the error due to
the residual scale dependence in the coefficient functions (as 
obtained after varying $m_b \leq \mu \leq 2 m_b$). 
If, instead of the results~(\ref{mi}), we take the values obtained 
by JLQCD~(\ref{oni}), the final number becomes 
$\left({\Delta \Gamma / \Gamma}\right)_{B_{s}} = 
6.8^{ +2.8}_{-3.4} \ \%$. 
So, the final values are numerically {\underline{small}} (much below the 
experimental limit). From eq.~(\ref{num}) we also see why it is so:   
$1/m_b$ corrections are very large and are of the sign opposite 
w.r.t. the second term, which would otherwise dominate eq.~(\ref{ours}). 
The matrix  elements that are present in $\delta_{1/m}$ are very hard to compute 
and it will take quite some time before the lattice results for 
$\delta_{1/m}$ appear. 

Finally, by using the same set of parameters and the results~(\ref{mi}), plus the value 
$f_{B_s} \approx 230(30)$~MeV~\cite{bernard}, from  eq.~(\ref{bbgln}) I obtain  
 a slightly higher central value, but a result totally
consistent with the above numbers, namely  
$\left({\Delta \Gamma/ \Gamma}\right)_{B_{s}} = 8.2^{ +3.1}_{-5.0}\ \%$.

Before closing this part, I would like to mention   
ref.~\cite{hurth} in which it has been argued that the width difference 
$\left({\Delta \Gamma / \Gamma}\right)_{B_{d}}$ might be esential for the
accurate determination of $\sin 2 \beta$ at the LHC. For details on the 
theoretical estimate of that quantity please see ref.~\cite{hurth}.  
Notice that they do not include the effects of 
the charm quark mass in the NLO correction to the coefficient functions. 
A main comment, however, is that like in the case of $\left({\Delta \Gamma / 
\Gamma}\right)_{B_{s}}$ also in this case it is highly important to 
get a  better control over $1/m_b$ corrections.

\section{RATIOS OF THE $B$-MESON LIFETIMES}

The hierarchy of the heavy meson lifetimes (for a given heavy quark), 
\bea
&&\tau(D^+) \ \gg \  \tau(D^0)\ \gg \ \tau(D_s)\;,\cr
&&\tau(B^+) \ \gtrsim \ \tau(B^0)\ \gtrsim \ \tau(B_s)\;,\nonumber 
\eea
can be explained by the effects of the spectator quark.
Theoretically formal way of expressing that, as in the previous section, 
is to perform the OPE, which reduces to identifying the local operators of 
${\cal O}(1/m_b^3)$. 
Obviously, the goal is to have an accurate theoretical determination 
of the ratios of the $B$-meson lifetimes, confront them to the precise experimental measurements, and 
therefore to test the underlying assumption of quark--hadron duality.
Although we are still a long way from that level of accuracy, the steady 
theoretical progress made over the last 10 years is rather 
encouraging.~\footnote{For selected reviews covering different 
aspects of the computation of these ratios, see refs.~\cite{neubert}.}

The spectator effects start showing up in OPE with the term 
$\propto 1/m_b^3$. Out of many $\Delta B=0$ local operators contributing 
at that order, only a few are expected to be relevant to the ratios 
$\tau(B_{u (s)})/\tau(B_{d})$. These have been identified in 
ref.~\cite{sachrajda}, and parametrized as follows:
\bea \label{eqA}
&& \langle    B_q\vert (\bar b q)_{V-A}
 (\bar q b)_{V-A} \vert  B_q\rangle = f_{B_q}^2 m_{B_q}^2 B_1
 (\mu)\,,\cr
&& \cr
&& \langle    B_q\vert \left(\bar b {\lambda^i\over 2} q\right)_{V-A}
 \left(\bar q {\lambda^i\over 2} b\right)_{V-A} \vert  B_q\rangle = f_{B_q}^2 m_{B_q}^2 \varepsilon_1
 (\mu)\,,\cr
&& \cr
&& \langle    B_q\vert (\bar b q)_{S-P}
 (\bar q b)_{S+P} \vert  B_q\rangle =f_{B_q}^2 m_{B_q}^2  B_2
 (\mu)\,,\cr
&& \cr
&& \langle    B_q\vert \left(\bar b {\lambda^i\over 2}  q\right)_{S-P}
 \left(\bar q {\lambda^i\over 2}  b\right)_{S+P} \vert  B_q\rangle = f_{B_q}^2 m_{B_q}^2  \varepsilon_2
 (\mu)\,.\eea
In the VSA, the colour singlet--singlet ($ss$) parameters are expected to be 
$B_{1}^{\rm VSA}=1$ and $B_{2}^{\rm VSA}=[m_{B_q}/(m_b + m_q)]^2\approx 1.5$, 
whereas the octet--octet ($oo$) ones are expected 
to give $\varepsilon_{1,2}^{\rm VSA}=0$.
The final expression for $\tau(B_{u})/\tau(B_{d})$ can be written as
\bea \label{rat}
{\tau(B^+)\over \tau(B^0) } &=& 1 \ + \ 16 \pi^2 {f_B^2 m_B\over m_b^3 c_3(m_b)} \
\biggl\{ \, G_1^{ss}(m_b)\ B_1(m_b)\ + \ G_1^{oo}(m_b) \ \varepsilon_1(m_b) \cr 
&&\cr
&&\hspace*{31mm}+\  G_2^{ss} (m_b)\ 
B_2(m_b)\ + \ 
G_2^{oo}(m_b)\ 
\varepsilon_2(m_b) + \bar \delta_{1/m_b} \biggr\}\,.
\eea
The main ingredients in this formula are:
\begin{itemize}
\item[$\circledast$] $16 \pi^2$ is the (``famous'') phase space enhancement 
of the spectator corrections ($\propto 1/m_b^3$);
\item[$\circledast$] $c_3(m_b)$ is the coefficient of the leading order term 
($\propto 1/m_b^0$) which survives the cancellation of the operators 
$\langle B_q\vert \bar bb\vert B_q\rangle$ in the ratio~(\ref{rat}). It 
consists of the phase space integrations in the total width of $B$-meson 
($b$-quark), plus the QCD radiative corrections. The NLO computation 
for $c_3(m_b)$ has been completed in ref.~\cite{bb}. 
An easier way to obtain this value (see~\cite{dunietz}) 
is to use the measured $b$-quark semileptonic branching fraction $B_{SL}^{\rm EXP.} = 
\Gamma(B\to X e\nu)/\Gamma_{TOT} = 10.6(3)\%$~\cite{lin}, and combine it 
with the theoretical expression for $\Gamma(B\to X e\nu)$~\cite{nir}. 
I obtain,
\bea
c_3(m_b) = 3.8(1)(3) \,,
\eea
where the last error comes from varying $m_c/m_b = 0.30 \pm 0.02$;
\item[$\circledast$] 
$G_{1,2}(\mu)$ are the functions describing the short distance QCD 
dynamics of $\Delta B=0$ operators. The situation with the computation 
of these functions is as follows. At LO in QCD, and by neglecting 
the charm quark mass ({\it i.e.} $z= m_c^2/m_b^2 = 0$), they were first 
computed in ref.~\cite{bigi2}. Inclusion of the finite charm-quark mass effects  
($z \neq 0$) was made in ref.~\cite{sachrajda}. This year's novelty is 
the computation of the NLO corrections. To get the final results, the authors of  
ref.~\cite{fred} keep  $z \neq 0$ in the LO term, 
and set $z=0$ in the NLO one.  
To better monitor the change in values for $G_{1,2}(\mu)$, 
I list all the functions needed in eq.~(\ref{rat}), both at LO and 
after including the NLO QCD corrections. 
{\TABLE[h]{
\begin{tabular}{|c|c|c|c|c|}
\cline{2-5}
\multicolumn{1}{l|}{}&  $G_1^{ss}(m_b)$  & $-G_1^{oo}(m_b)$ & $G_2^{ss}(m_b)$  & $G_2^{oo}(m_b)$\\
\hline \hline
LO($z=0$) &      0.19 &   10.03 &          0.06 &        2.43\\
LO($z=0.3^2$) &  0.16 &    8.40 
&          0.06  &        2.37 \\ \hline 
NLO($z=0$) & {\bf 0.52 } & {\bf 9.60}  &{\bf 0.03}  & {\bf 1.86} \\
NLO($z=0.3^2$ and $z=0$) & {\bf 0.55  } 
& {\bf 8.08 }  
& {\bf 0.03 }  
& {\bf 1.80 } \\
\hline 
\end{tabular}
\caption{\sl The values of the functions $G_i(m_b)$ 
appearing in eq.~(\ref{rat}) as obtained in ref.~\cite{fred}. 
The corresponding numbers relevant to the ratio $\tau(B_{s})/\tau(B_{d})$
can be found in that reference.}
\label{tab1}}}

Qualitatively, as in the case of $(\Delta \Gamma/\Gamma)_{B_s}$, the authors of
ref~\cite{fred} explicitly verify the 
infrared 
safety of $G^{\rm NLO}_{1,2}(\mu)$. However, they do not estimate the    
residual scale dependence of $G_{1,2}(m_b)$. For the phenomenological 
considerations I will add (hopefully) conservative $\pm 20$\% of that error.
Concerning the change in value for each of the functions $G_{1,2}(m_b)$, 
we see from table~\ref{tab1} that they all receive rather moderate radiative 
corrections except for the ``dramatic'' case of $G_1^{ss}(m_b)$, whose central value changes as much as 
\bea
{  G_1^{ss\ \rm NLO} (m_b) - G_1^{ss\ \rm LO}(m_b)\over 
G_1^{ss\ \rm LO}(m_b) } \simeq 2.5\; .
\eea

\item[$\circledast$] $\bar \delta_{1/m_b}$ stands for the neglected 
terms $\propto 1/m_b^3$ which are not enhanced by ``$16 \pi^2$", 
and for the terms in OPE that are $\propto 1/m_b^4$. A discussion of the former
has been made in ref.~\cite{pirjol}, while for the latter no research has been
made to date. It would be nice to follow the lines of 
ref.~\cite{dunietz} and check whether the VSA indicates 
 $\bar \delta_{1/m_b}$ to be small or large.

\item[$\circledast$] Until this year, there was only 
one lattice study of the matrix elements~(\ref{eqA}), 
and that one was made in the static limit of HQET~\cite{massimo} (see 
also ref.~\cite{flynn}). 
This year, the first lattice QCD computation of $\Delta B=0$ operators 
has been performed~\cite{damir2}, the main features of which will be 
explained in the next subsection. Besides lattice simulations, also the QCD sum 
rule methodology was employed in HQET to estimate the wanted bag 
parameters~\cite{chung}. The compendium of the present results looks 
as follows:
\bea \label{params}
&&\hspace*{-0.5mm}{\underline{\rm Sum\ rules\ (HQET)}}~{\rm \cite{chung}}\hspace*{14mm}
{\underline{\rm Lattice\ HQET}}~{\rm \cite{massimo}}\hspace*{16mm}
{\underline{\rm Lattice\ QCD}}~{\rm \cite{damir2}}
  \nonumber \\
&&B_1(m_b) = 1.01(1)\,\hspace*{22.5mm}B_1(m_b) =1.06(8)\, \hspace*{17.5mm}B_1(m_b)
=1.10(13)\left({}^{+.10}_{-.21}\right)\,\nonumber \\
&&B_2(m_b) = 0.99(1)\,\hspace*{2.25cm}B_2(m_b) =1.01(6)\, 
\hspace*{1.75cm}B_2(m_b) = 0.79(5)(9)\,\nonumber \\
&&\varepsilon_1(m_b) = -0.08(2)\,\hspace*{2.1cm}\varepsilon_1(m_b) = -0.01(3)\,
\hspace*{1.55cm}\varepsilon_1(m_b) = -0.02(2)\left({}^{+.01}_{-.00}\right)\, \nonumber \\
&&\varepsilon_2(m_b) = -0.01(3)\,\hspace*{2.1cm}\varepsilon_2(m_b) = -0.01(2)\, 
\hspace*{1.55cm}\varepsilon_2(m_b) = 0.03(1)\left({}^{+.01}_{-.00}\right)\,\nonumber \\
\eea
\end{itemize}

\FIGURE{
\epsfig{file=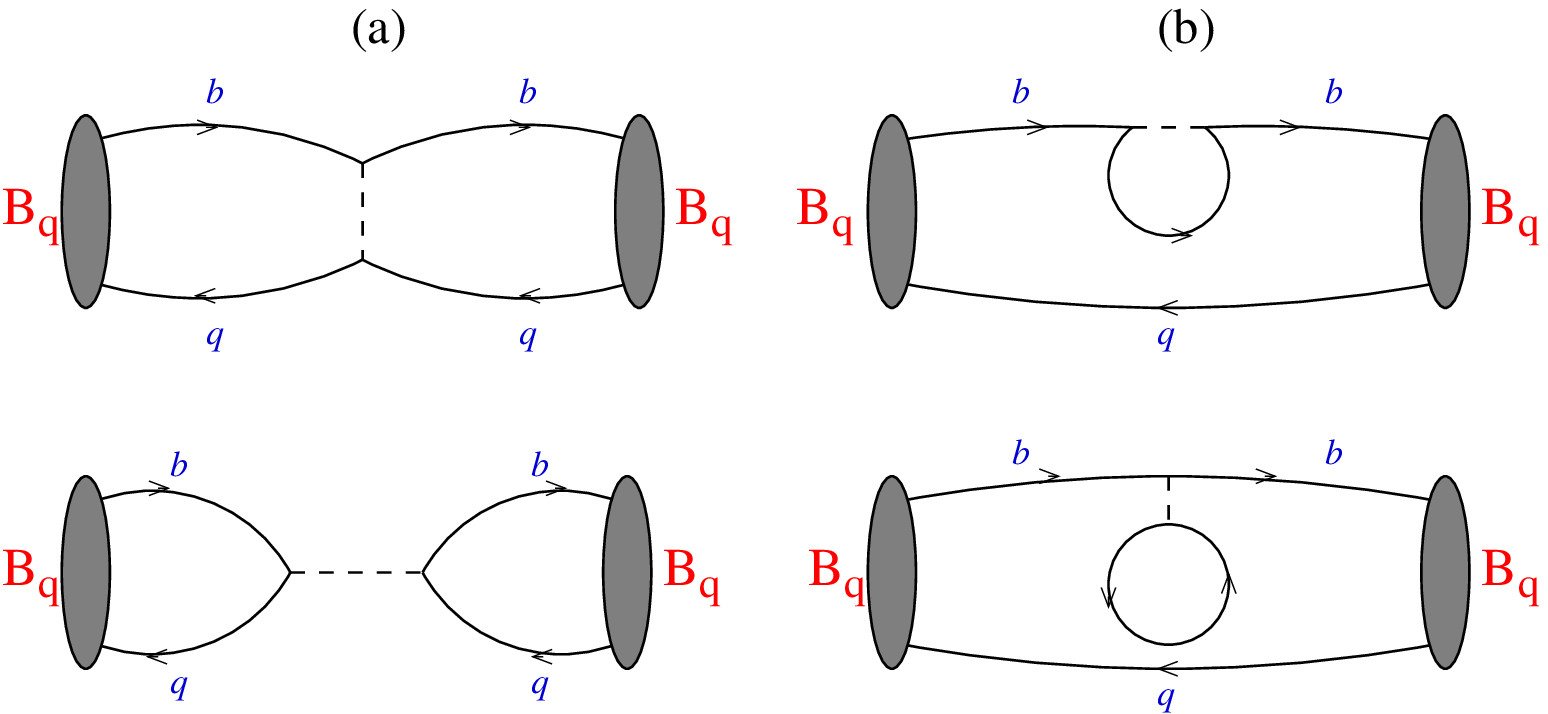,width=0.98\linewidth}
\caption{\sl Graphs (a) are computed on the lattice. Graphs (b) are not.
Dashed lines denote a $\Delta B=0$ four-quark operator~(\ref{eqA}).}
\label{fig3}} 
Before updating the values for $\tau (B_{u,s})/\tau (B_{d})$, I 
stop here to give a few details concerning the lattice QCD 
computation of $\Delta B=0$ operators. 
The reader not interested in lattice QCD {\it ``cuisine''} may skip the next 
subsection.

\subsection{Lattice QCD estimate of $B_{1,2}(m_b)$ and $\varepsilon_{1,2}(m_b)$}

Since the paper containing details about this computation  
has not been released~\cite{damir2}, 
I feel it is fair to the lattice community (and wider) to explain a few elements involved 
in this computation. We employ the  
$24^3 \times 48$ lattice at $\beta = 6.2$ ({\it i.e.} $a^{-1}=2.7(1)$~GeV) and
use the Wilson fermions to compute the diagrams shown in
fig.~\ref{fig3}(a). We have three values of the heavy and three
values of the light quark masses. For easier orientation, our heavies are 
around the charm quark mass, while the lights are around the strange one. 
It is convenient to redefine the bag parameters for the operators as 
\bea \label{redef}
B_2 \to \overline B_2 \cdot {m_{B_q}^2\over (m_b + m_q)^2} \;, \quad \quad
\varepsilon_2 \to \overline \varepsilon_2  \cdot  {m_{B_q}^2\over (m_b + m_q)^2} \;,
\eea 
so that
$\overline B_2^{VSA} =1$. To subtract the spurious mixings with other 6 
dimension-six
operators from the
bare operators (peculiarity of the lattice computation with Wilson fermions), 
and to match them with the continuum ones, renormalized in the
RI/MOM scheme, we used the 1-loop (boosted) perturbative expressions of
ref.~\cite{lanl}. Since the NLO coefficients in this matching procedure are
quite large it is desirable to renormalize the operators 
non-perturbatively and check 
the impact on the values presented in eq.~(\ref{params}).
That work is in progress. The bag parameters ($B_1$, $\overline B_2$, 
$\varepsilon_1$, $\overline \varepsilon_2$) are extracted in the usual way,
that is from the suitable ratios of the three-point and two-point correlation functions. 
We convert the extracted values from RI/MOM to the $\overline {\rm MS}$(NDR) renormalization  
scheme of ref.~\cite{buras}, to combine them with the coefficient 
functions $G_{1,2}(m_b)$ from table~\ref{tab1}.

The extrapolation of all the bag parameters in the light quark mass, 
to the physical $up/down$ quark, has been done linearly. Finally, these results 
are extrapolated in the inverse heavy meson mass, as shown in
fig.~\ref{fig3}. We matched the leading order 
anomalous dimension matrices in QCD with the ones in HQET 
($\Phi(m_P)$ in fig.~\ref{fig4}), in such a way that the 
extrapolated quantity $\Phi(m_B)$ leads directly to the desired bag parameters 
in QCD. 
\FIGURE{
\epsfig{file=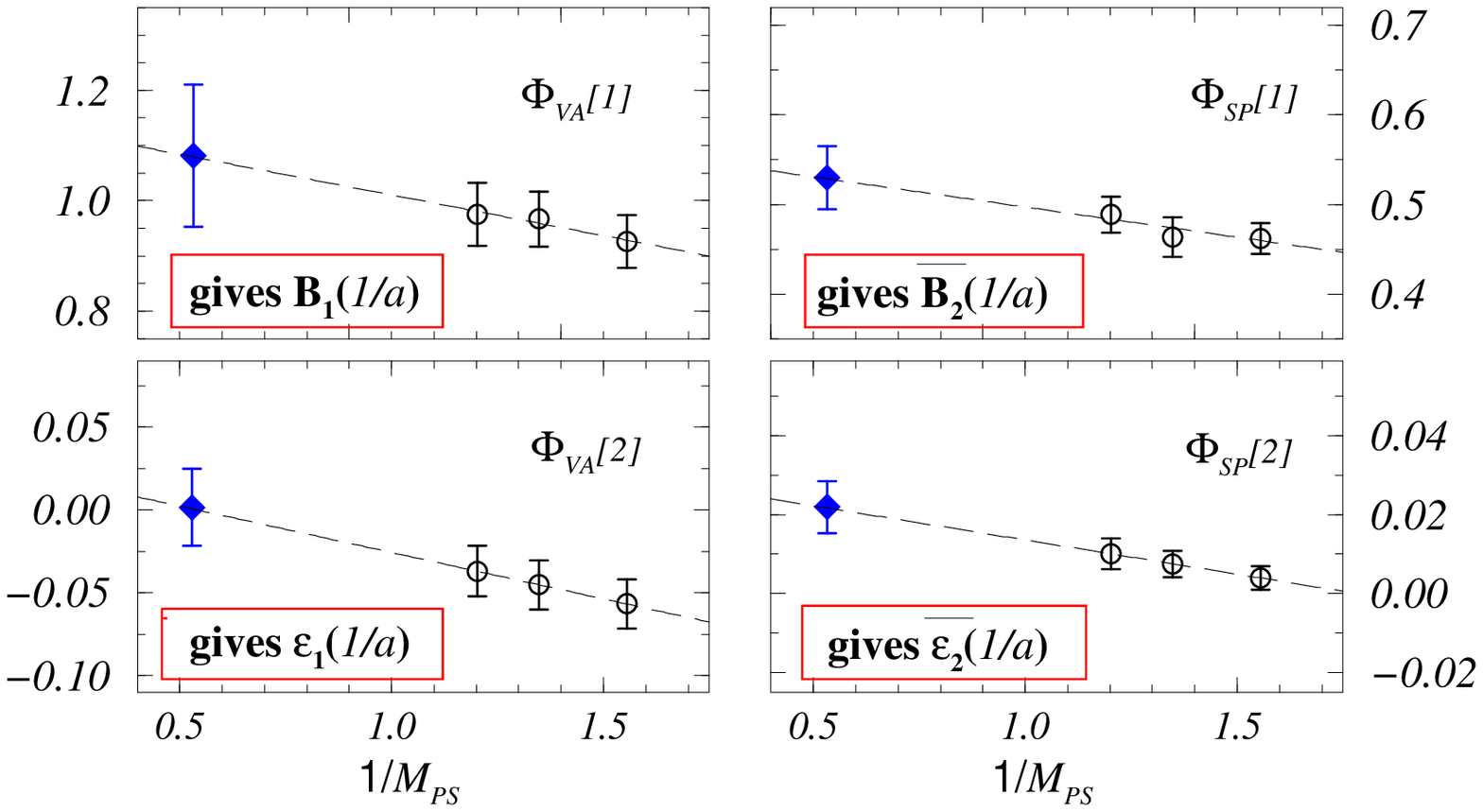,width=0.94\linewidth}
\caption{\sl Extrapolation of the bag parameters of $\Delta B=0$ operators 
in inverse heavy meson mass (displayed in lattice units) from the masses
accessible from our lattice study (empty symbols) to the $B$-meson mass 
(filled symbols). The plotted numbers correspond to the $\overline {\rm MS}$(NDR) scheme 
and $\mu = 1/a = 2.7(1)$~GeV.}
\label{fig4}} 
To get the numbers given in eq.~(\ref{params}), 
we transform $ \overline B_2 \to B_2$ and $ \overline \varepsilon_2 \to 
\varepsilon_2$ as indicated  
in eq.~(\ref{redef}). 
Many sources of systematic uncertainties are included
in the second error in eq.~(\ref{params}).
Please note that the parameter $B_2$ is quite far from its VSA value
 (our $B_2(m_b)\approx 0.8(1)$, whereas $B_2^{VSA}\approx 1.5$).

\subsection{Lifetime ratios: final touch}
We are now ready to combine all the above results and update the theoretical 
prediction concerning the ratio of the heavy meson lifetimes. By using 
$f_B=0.198(30)$~GeV~\cite{bernard}, I finally obtain
\bea
{\tau(B^+)\over \tau(B^0) } \; =\; 1.067(27)\; \simeq \; 1.07(3) \;.
\eea
If we take
for the bag parameters the results obtained from the lattice HQET (without
the appropriate NLO matching to QCD!), we get ${\tau(B^+)/ \tau(B^0) } = 1.06(3)$.

As for the ratio ${\tau(B_s)/\tau(B_d) }$, I combine the NLO-corrected values 
for the coefficient functions~\cite{fred} with the bag parameters obtained 
in lattice QCD~(\ref{params}), to arrive at
\bea
{\tau(B_s)\over \tau(B_d) }\, =\, 0.998(15) \, \simeq \, 1.00(2) \;. 
\eea
This result remains unchanged when the bag parameters are replaced by those determined
from the lattice HQET.

\subsection{Cum grano salis}

In the computation of the bag parameters, we did not include the penguin-like 
contractions shown in fig.~\ref{fig3}(b). The renormalization of such a diagram 
in QCD is very difficult, because the $\Delta B=0$ dimension-six operators may mix with 
the lower dimensional ones ({\it e.g.} $\bar b b$); we thus have to first make 
a power subtraction of such mixings~\footnote{This issue is the familiar problem 
present in the lattice computation of $\Delta I=1/2$ amplitude in the $K\to \pi \pi$ decay.}, 
followed by the (standard) multiplicative renormalization. Up to now, there is no method 
allowing such a computation non-perturbatively. 
Notice that in ref.~\cite{massimo}, the contractions from fig.~\ref{fig3}(b) were omitted too.

A similar problem appears in perturbation theory when one computes the coefficient 
functions~$G^{\rm NLO}_{1,2}(m_b)$ by including penguins similar to those 
shown in fig.~\ref{fig3}(b). The situation becomes even worse because, 
for dimensional reasons, the mixing with {\it e.g.} $\bar b b$ is $\propto \alpha_s(m_b) 
m_b^3 \bar b b$, so that the counting in the OPE breaks down. 
The way out is to match the QCD with the HQET, where the heavy quark is 
completely  integrated out and the counting in $1/m_b$ is guaranteed from 
the first principles. In addition, no spurious mixings with lower 
dimensional operators appear in (dimensionally regularized) 
perturbation theory. A complete discussion of that point can be 
found in ref.~\cite{fred}.

After having made this important remark, one also needs to give a rationale for neglecting 
those (in)famous contractions. The argument, as presented in ref.~\cite{sachrajda}, 
says that since these penguin-like contractions do not directly include the spectator quark 
effects, their presence in the ratio ${\tau(B_u)/ \tau(B_d)}$ can be safely neglected, 
owing to the isospin symmetry. In other words, a given $\Delta B=0$  operator in 
fig.~\ref{fig3}(b) does not feel a switch $u\leftrightarrow d$ on the spectator line, 
and since such contractions appear as a difference in the ratio ${\tau(B_u)/ \tau(B_d)}$, 
their contributions cancel.
The argument holds also for ${\tau(B_s)/ \tau(B_d)}$, but in this case the assumption is 
somewhat stronger since the SU(3) light flavour symmetry has to be invoked.

Of course, this argument is not valid if one needs to estimate a combination 
$B_1(\mu) - B_2(\mu)$, 
which is relevant for the precision extraction of $\vert V_{ub}\vert$ 
from the inclusive semileptonic $B\to X_u \ell \nu_\ell$ decay~\cite{misha}. 
For this study, the inclusion of the penguin-like contractions of fig.~\ref{fig3}(b) is
mandatory.

\section{Summary and prospects}

After several years of exciting research to reduce the theoretical 
uncertainty on $(\Delta \Gamma/\Gamma)_{B_s}$, the leading term in OPE 
for this quantity is in good shape: 
NLO perturbative corrections and quite reliable estimates of 
the matrix elements,  
obtained from new lattice studies, are available. However, 
a rough estimate of the subleading ($1/m_b$) 
corrections in the OPE indicates that such corrections, to a large extent,  
wash out the effect of the leading terms: the corrections are large and of 
opposite sign. Therefore, as of now, it does not seem reasonable  to 
test the Standard Model (or to expect to 
see the signal of physics beyond the Standard Model) from this quantity. 
It is, in fact, necessary to improve the theoretical predictions by 
taming the dimension-seven operators (the ones that enter with $1/m_b$ corrections). 
From the present theoretical situation I conclude that 
\bea
\left({\Delta \Gamma \over \Gamma}\right)_{B_s}\ =\ \left( \ 7\ \pm \ 4 \ \right)\ \%\,.
\eea

This year, a further theoretical improvement in the lifetime ratios of the $B$-mesons 
has been made. QCD radiative corrections to the coefficient functions are now being 
calculated, and the first QCD computation of the bag parameters performed (which is 
complementary to the earlier lattice HQET results). 
Lattice predictions will certainly improve in many respects ({\it e.g.} non-perturbative 
renormalization will be carried out, the penguin-like contractions are likely to
 be included in the 
HQET lattice studies, unquenched simulations in HQET will become feasible,\dots)
 Note that the operators that give rise to $1/m_b$ corrections to the non-spectator 
effects in  ${\tau(B_{u,s})/ \tau(B_d)}$ are yet to be identified and their effects 
estimated. Such a study would be very welcome. 
From the present theoretical information, for the ratios of the $B$-meson 
lifetimes I 
obtain
\bea
{\tau(B^+)\over \tau(B^0) } \ = \ 1.07(3) \;,\hspace*{12mm} 
{\tau(B_s)\over \tau(B_d) }\, =\, 1.00(2) \;. 
\eea

\vspace*{2.5cm}

\subsection*{Acknowledgement}
It is a pleasure to thank my friends and collaborators of refs.~\cite{ape},
\cite{damir} and also those of ref.~\cite{fred}, for 
sharing their insights in topics covered by this talk. 
Communication with S.~Ryan, N.~Yamada and the authors of 
ref.~\cite{massimo} is kindly acknowledged.

\end{document}